\documentclass[12pt]{article}
\usepackage{graphicx,amsmath,amssymb}

\newcommand{\beq}{\begin{equation}}
\newcommand{\eeq}{\end{equation}}

\def\bmp#1{\begin{minipage}{#1\textwidth}}
\def\emp{\end{minipage}}

\newcommand\pubdate{\today}

\def\Cincy{Department of Physics, University of Cincinnati, Cincinnati, Ohio 45221,USA}
\def\LjubljanaFMF{Faculty of Mathematics and Physics, University of Ljubljana,\\
 Jadranska 19, 1000 Ljubljana, Slovenia }
\def\LjubljanaIJS{Josef Stefan Institute, Jamova 39, 1000 Ljubljana, Slovenia}
\def\SISSA{SISSA, Via Bonomea 265, I34136 Trieste, Italy}

\def\Title#1{\begin{center} {\Large #1 } \end{center}}
\def\Author#1{\begin{center}{ \sc #1} \end{center}}
\def\Address#1{\begin{center}{ \it #1} \end{center}}

\newcommand\pubblock{\rightline{\begin{tabular}{l}
         \pubdate  \end{tabular}}}
\newenvironment{Abstract}{\begin{quotation}  }{\end{quotation}}
\newenvironment{Presented}{\begin{quotation} \begin{center} 
             PRESENTED AT\end{center}\bigskip 
      \begin{center}\begin{large}}{\end{large}\end{center} \end{quotation}}
\def\Acknowledgements{\bigskip  \bigskip \begin{center} \begin{large}
             \bf ACKNOWLEDGEMENTS \end{large}\end{center}}

\def\beq{\begin{equation}}
\def\eeq#1{\label{#1}\end{equation}}
\def\eeqn{\end{equation}}

\def\beqa{\begin{eqnarray}}
\def\eeqa#1{\label{#1}\end{eqnarray}}
\def\eeqan{\end{eqnarray}}

\let\bar=\overbar

\def\Dslash{\not{\hbox{\kern-4pt $D$}}}
\def\dslash{\not{\hbox{\kern-2pt $\del$}}}

\def\msb{{\bar{\ssstyle M \kern -1pt S}}}

\begin{document}
\begin{titlepage}
\pubblock

\vfill
\Title{\bf The case for measuring gamma precisely}
\vfill
\Author{ Jure Zupan}
\Address{
\Cincy\\
\LjubljanaFMF\\
\LjubljanaIJS\\
\SISSA}
\vfill
\begin{Abstract}
We first review the methods for determining $\gamma$ from $B\to D K$ decays that appeared after CKM 2008. We then discuss the theoretical errors in $\gamma$ extraction. The errors due to neglected $D-\bar D$ and $B_{d,s}-\bar B_{d,s}$ mixing can be avoided by including their effects in the fits. The ultimate theoretical error is then given by electroweak corrections that we estimate to give a shift $\delta \gamma/\gamma\sim {\mathcal O}(10^{-6})$.  
\end{Abstract}
\vfill

\begin{Presented}
The 6th International Workshop on the CKM Unitarity Triangle, University of Warwick, UK, 6-10 September 2010
\end{Presented}
\vfill
\end{titlepage}
\def\thefootnote{\fnsymbol{footnote}}
\setcounter{footnote}{0}

\section{Introduction}
\begin{figure}
\begin{center}
${}$\\[-30mm]
\fbox{
\bmp{0.35}
{ a)} ${}$\\[-10mm]
\includegraphics[width=4.5cm]{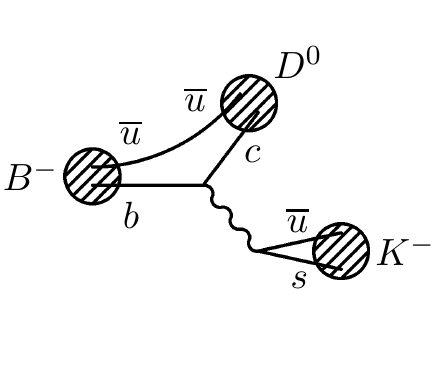}\\[-10mm]
\centerline{{\Huge $\Updownarrow$}}
${}$\\[-15mm]
\includegraphics[width=4.cm]{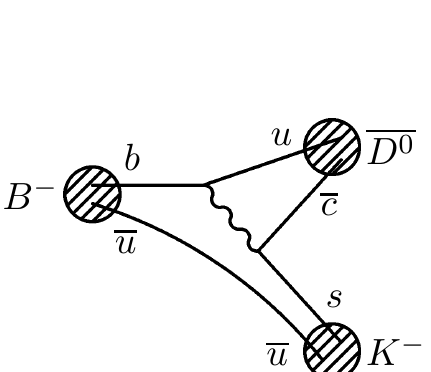}
\emp
}
~~~~~~~
\bmp{0.50}
${}$\\[50mm]
~~~~{b)}\\[-20mm]
${}$\hspace{-7mm}\includegraphics[width=9cm]{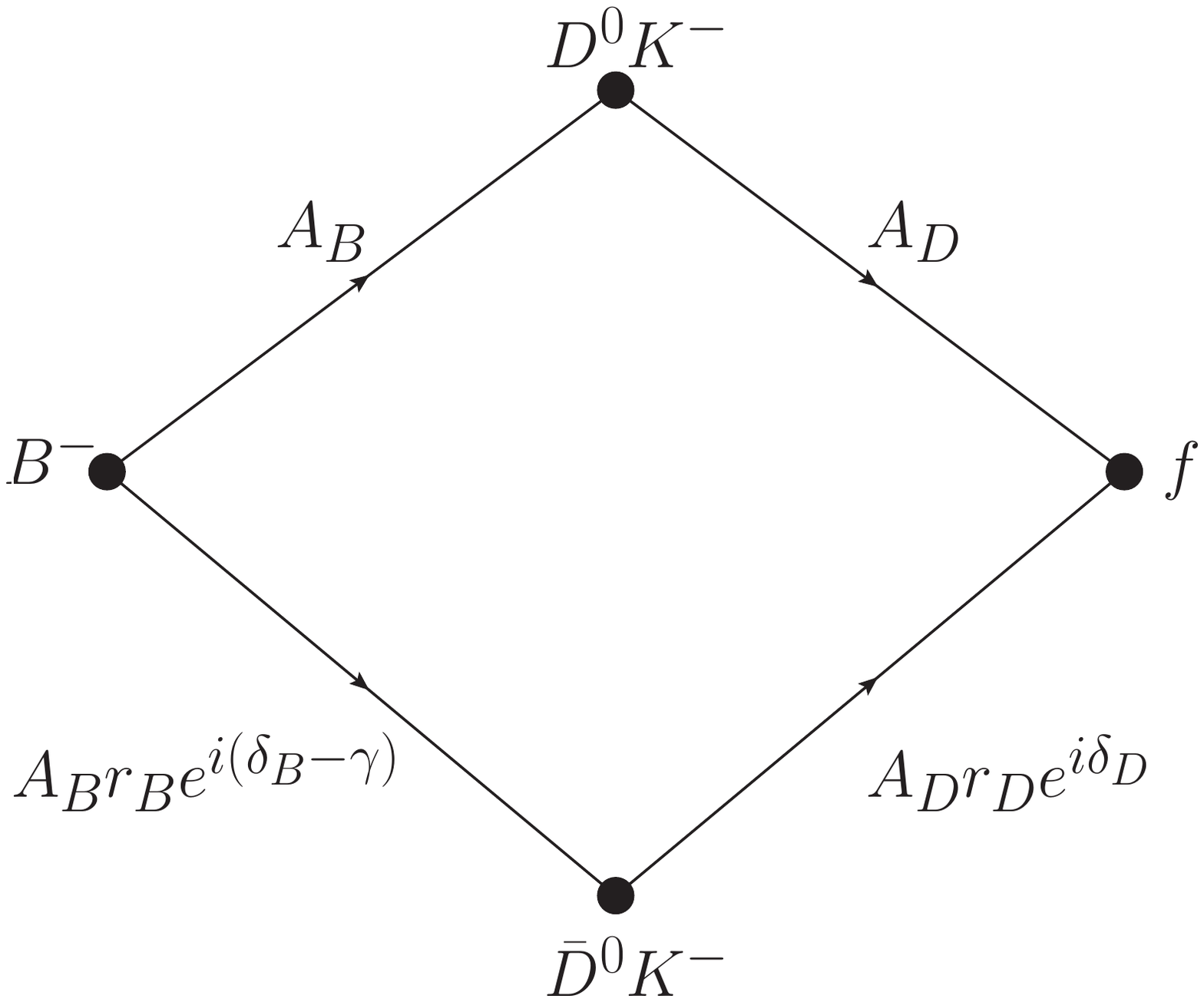}
\emp
\end{center}
${}$\\[-60mm]
\caption{\sl\small The color allowed and color suppressed amplitudes interfering  in $B^\pm\to DK^\pm$ decays (left), with $D^0$ and $\bar D^0$ decaying to a common final state $f$ (right).}\label{decaychainGraph}
\end{figure}

It is exactly 20 years ago since  Gronau in two papers with London and with Wyler pointed out that the interference between $b\to c \bar{u} s$ and $b\to u \bar{c} s$ transitions can be used to obtain the CKM weak phase $\gamma$ \cite{Gronau:1990ra,Gronau:1991dp}.  The interference between $B\to D K$ and $B\to \bar{D} K$ arises, if both $D$ and $\bar{D}$ decay to a common final state $f$ (cf. Fig. \ref{decaychainGraph}, which also defines our notation). 
What makes this approach very powerful is that 
there are no penguin contributions and therefore almost no theoretical uncertainties, with all the
hadronic unknowns in principle obtainable from experiment.

Different methods can be grouped according to the choice of the final state $f$, which can be (i)
a CP- eigenstate (e.g. $K_S \pi^0$) \cite{Gronau:1991dp}, (ii)
a flavor state (e.g. $K^+\pi^-$) \cite{Atwood:1996ci}, (iii)
a singly Cabibbo suppressed (e.g. $K^{*+} K^-$) \cite{Grossman:2002aq} or (iv) 
a many-body final state (e.g. $K_S\pi^+\pi^-$) \cite{Giri:2003ty}. There are also other extensions:
many body $B$ final states (e.g. $B^+\to D K^+\pi^0$, etc) \cite{Aleksan:2002mh,Gershon:2009qr,Gershon:2008pe, Gershon:2009qc}, 
 $D^{0*}$ in addition to $D^0$ (taking care of a sign flip in the use of $D^*$ \cite{Bondar:2004bi}), self tagging $D^{0**}$ \cite{Sinha:2004ct,Gershon:2008pe} or neutral $B$ decays (time dependent and time-integrated as well as $B^0$ and $B_s$) can be used \cite{Gronau:2004gt,Kayser:1999bu}. In this write-up we will review only the methods to extract $\gamma$ that appeared after the CKM2008 conference (for older methods see, e.g. \cite{Zupan:2007zz}). In the remainder we will then focus on the theoretical errors. 

\section{New method(s) since CKM2008}
When we talk about a ``method'' to extract $\gamma$ this really just denotes a subset of final states allowing for the extraction of this weak phase. This may appear as  a not very sensible way of approaching the problem, since combining as many final states as possible increases our sensitivity to $\gamma$. However, splitting them in terms of "methods" can also be benificial. It offers a possibility to check for systematics or even search for New Physics. 

Most of the methods date to the 1990s and early 2000s \cite{Zupan:2007zz}, with some development also in the last two years. Since 2008 a new set of methods using multibody $B^0\to DK^+\pi^-$ modes has been worked out in Refs. \cite{Gershon:2009qr,Gershon:2008pe}. This multibody $B$ decay Dalitz plot features a flavor specific resonant decay $D_2^{*-}(2460)\to \bar D^0\pi^-$. The interference with the other  resonances in the $DK^+\pi^-$ Dalitz plot, for instance with the $B^0\to D K^{*0}(1430)$ or $B^0\to D K^{*0}(892)$, gives the sensitivity to extract $\gamma$. There are still many choices for $D\to f$ decays that one can make. 

The important thing to note is, that because $D_2^{*-}(2460)\to \bar D^0\pi^-$ is flavor specific and interferes with the other resonances, one needs fewer $D\to f$ final states then if $B$ had been decaying in a two-body final state. In particular, the equivalent of GLW method \cite{Gronau:1990ra,Gronau:1991dp} does not need a CP-odd $D\to K_S\pi^0$ decay, which would be difficult for LHCb. 
Compared to quasi two-body $B^0\to DK^{*0}$ decays one thus achieves at least 50\% better sensitivity to $\gamma$ \cite{Gershon:2009qc}. In Ref. \cite{Gershon:2009qr} the model independent method \cite{Giri:2003ty} was demonstrated for the double Dalitz plot analysis $B^0\to DK^+\pi^-\to (K_S\pi^+\pi^-)_DK^+\pi^-$.  Using reasonable models for the poorly known $B^0\to DK^+\pi^-$ Dalitz plot the authors estimate that 20 annual yields of LHCb would lead to an ${\mathcal O}(1^\circ)$ error.

All these methods are statistics dominated at present, so it is natural to combine them into a global fit to $\gamma$. At the end, we are really interested in $\gamma$. And for this one gains more then just statistics when combining different channels. For different $D$ decays in $B\to (f)_D K$ the $B$ system parameters are common and this leads to additional gain. For $N_D$ different $D$ decay channels and $N_B$ different $B$ decay modes one has $\sim N_D N_B$ measurements, while there are only $\sim N_D +N_B$ unknowns. Because of this quadratic increase of measurements compared to the linear rise of unknowns, the combined analysis will always win over the separate methods in the precision of extracted $\gamma$ (or even over combining later the results on $\gamma$ from different methods). 

However, splitting the data into subsets provides us with a way to  check for systematics or search for New Physics. In fact, the way $\gamma$ extraction is implemented at Belle and BaBar now, it already is an automatic test of NP contributions in decay amplitudes. Consider the case where the decay amplitudes get modified by an extra contribution with a new strong phase $\delta_B'$ and a weak phase $\gamma'$
\begin{equation}
\begin{split}
A(B^\pm \to f_D K^\pm)&\propto 1+r_D e^{i\delta_D}(r_B e^{i(\delta_B\pm\gamma)}+r_B' e^{i(\delta_B'\pm\gamma')}).\label{fBarDec}
\end{split}
\end{equation}
This means that for $B^+$ and $B^-$ we have different $r_B$
\begin{equation}
r_{B^+}\to |r_Be^{i(\delta_B+\gamma)}+r_B'e^{i(\delta_B'+\gamma')}|, \qquad r_{B^-}\to |r_Be^{i(\delta_B-\gamma)}+r_B'e^{i(\delta_B'-\gamma')}|.
\end{equation}
Having $r_{B^-}\ne r_{B^+}$ would signal a NP contribution to the $B\to DK$ amplitude. The nice thing about this check is that BaBar and Belle already measure all the necessary ingredients, since they have mapped the problem of measuring $\gamma$ to the problem of measuring $x_{\pm}=r_B\cos(\gamma\pm\delta_B)$, $y_{\pm}=\pm r_B\sin(\gamma\pm\delta_B)$. The sign of NP would then be 
\begin{equation}
{x_+^2+y_+^2}\ne {x_-^2+y_-^2}.
\end{equation}
Of course, even if $x_+^2+y_+^2= x_-^2+y_-^2$, there can be a shift in the extracted $\gamma$ from NP or systematics. A test that goes beyond this is that $\gamma$ from $B^\pm\to DK^\pm$, $B^\pm\to DK^{*\pm}$, $B^\pm\to D^*K^\pm$, $B^0\to DK^0$, ..., all coincide.

\section{Theory errors on extracting $\gamma$}
The determination of $\gamma$ from $B\to DK$ decays is theoretically extremely clean since these are pure tree decays. In this section we estimate the theoretical errors inherent to the method, and thus the ultimate precision one can hope to achieve.
The largest uncertainty is due to the $D-\bar D$ mixing, which we neglected so far, because it is still very small. 
In SM $D-\bar D$ mixing is CP conserving to a very good approximation, with the mixing phase $\theta\sim O(10^{-4})$, while the effect on $\gamma$ is $O(x^2, y^2)$ \cite{Grossman:2005rp}. Here $
x \equiv {\Delta m_D \over \Gamma_D} , y \equiv {\Delta \Gamma_D \over 2\Gamma_D} 
$, with $x\sim y\sim O(10^{-2})$.

If the $D$ decay information is coming from flavor tagged $D$ decays (i.e. from $D^*\to D\pi)$ then CP conserving $D-\bar D$ mixing only affects the interference term between $A_B$ and $A_Br_B$ amplitudes \cite{Grossman:2005rp}. It changes the relative strong phase $\delta_D$ to a time averaged effective strong phase $\langle \delta _D\rangle$. Since it is fit from data anyway,  this does not affect $\gamma$ extraction. The mixing also dilutes the interference, which results in the shift on $\gamma$ of ${\mathcal O}(x^2/r_D^2,y^2/r_D^2)$. The largest shift is then for  doubly Cabibbo suppressed $D$ decays. But even then $\Delta \gamma\lesssim 1^\circ$.
In model independent Dalitz plot analysis no changes are needed, if all information comes from $B\to DK$ decays, since in the method one already fits also for the dillution of the interference term.

The effect is potentially larger, if $D$ decay information is coming from entangled $\psi(3770\to D\bar D)$ decays, i.e. if one uses information from CLEO and BES-III \cite{Bondar:2010qs}. The difference is that in entangled $D$ decays the time integration interval is $t\in (-\infty, \infty)$, while in $B\to D K$ decays the time integration interval is $t\in [0,\infty)$. As a result the shift in $\gamma$ now starts at first order in mixing and is thus linear in $x_D, y_D$. Even so, the shift in $\gamma$ is still small, $\Delta \gamma \leqslant 2.9^\circ$. It is even smaller, $\Delta \gamma \leqslant 0.2^\circ$, if $|A_D|^2$ is measured from $D^*\to D\pi$ and the entangled decays are used only for the determination of the interference terms. Most importantly, by simply modifying the equations the $D-\bar D$ mixing effects can be included exactly, if $x_D, y_D$ are precisely measured \cite{Silva:1999bd}. Similarly, for $\gamma$ extraction from untagged $B_s\to D \phi$ decays the inclusion of $\Delta \Gamma_s$ can be important and can be achieved once $\Delta \Gamma_s$ is well measured \cite{Gronau:2007bh}.

\begin{figure}
\begin{center}
~~~~
\bmp{0.45}
{ a)}\\
\includegraphics[width=4cm]{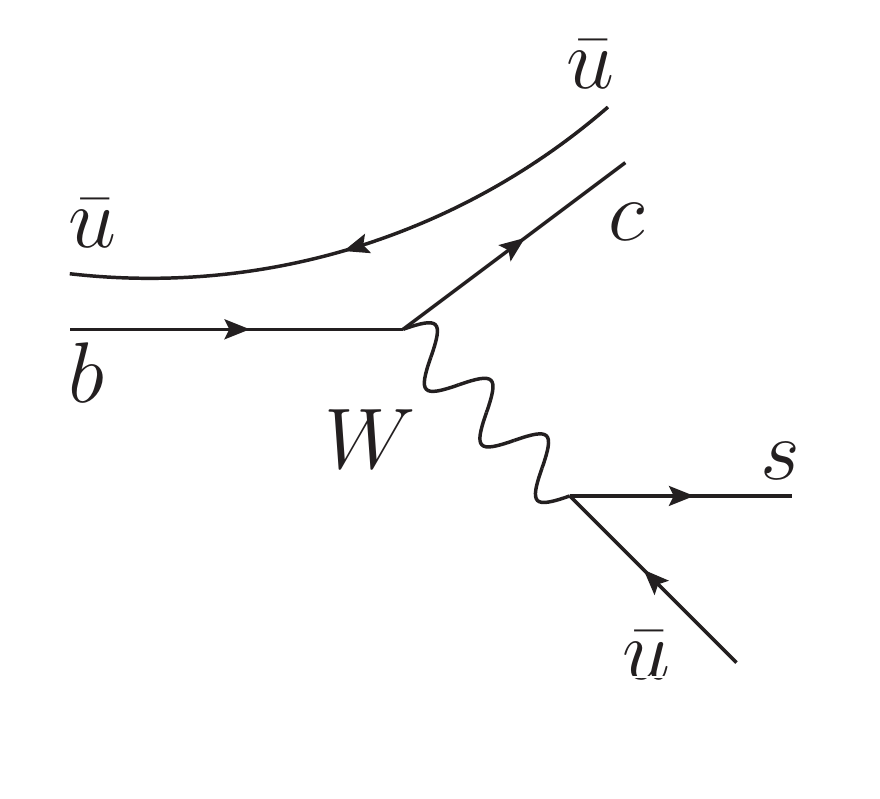}
\emp
~~
\bmp{0.45}
{ b)}\\
\includegraphics[width=5cm]{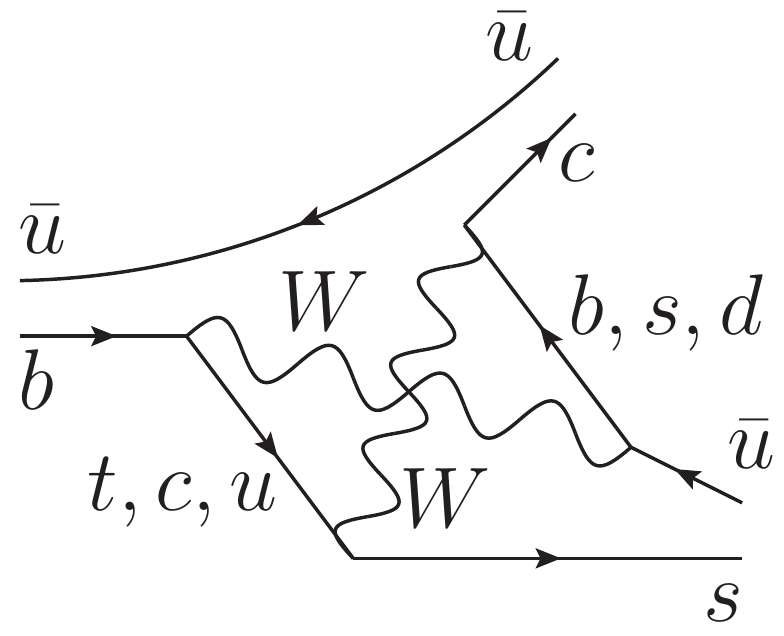}
\emp
\end{center}
${}$\\[-16mm]
\caption{\sl\small A box diagram electroweak corrections (b) with a different CKM structure than the leading weak decay amplitude (a).}\label{higherEW}
\end{figure}

The remaining SM theory error then comes from higher electroweak corrections. This error is not so easy to get rid of using just the experimental information and may well represent the ultimate precision of the approach. Not all electroweak corrections matter though - the important ones are the corrections that change the CKM structure. For instance, vertex corrections and $Z$ exchanges do not affect the $\gamma$ extraction. The corrections from box diagrams, Fig. \ref{higherEW}, on the other hand, do 
carry a different weak phase than the LO. For $(\bar c b)_L(\bar s u)_L$ operator the relevant corrections carry a weak phase. The leading correction is 
\begin{equation}
\sim \frac{g^2}{16\pi^2}V_{cb} V_{cs}^* V_{ub}^* V_{cb} \frac{m_b^2}{m_W^2}\big[A_B/(V_{cb} V_{us}^*)\big]\sim \frac{g^2}{16\pi^2}\lambda^4 \frac{m_b^2}{m_W^2}A_B.
\end{equation}
For $(\bar u b)_L(\bar s c)_L$ operator the relevant contributions have a weak phase different from $\gamma$. The leading correction is (note that terms with intermediate top are power suppressed)
\begin{equation}
\sim \frac{g^2}{16\pi^2}V_{cb} V_{cs}^* V_{cs}^* V_{us} \frac{m_c^2}{m_W^2}\big[A_B/(V_{cb} V_{us}^*)\big]\sim \frac{g^2}{16\pi^2}\lambda^4 \frac{m_c^2}{m_W^2}A_B.
\end{equation}
This contribution dominates the irreducible theory error on $\gamma$ with $\delta \gamma/\gamma\sim {\mathcal O}(10^{-6})$. 

\begin{table}
\begin{tabular}{cccc}
\hline
\hline
Probe & $\Lambda_{NP}$ for (N)MFV NP &$\Lambda_{NP}$ for gen. FV NP & $B\bar B$ pairs\\ 
\hline
\hline
$\gamma$ from $B\to DK^{1)}$ & $\Lambda\sim {\mathcal O}(10^2$ TeV)   & $\Lambda\sim {\mathcal O}(10^3$ TeV) & $\sim 10^{18}$ \\
$B\to \tau \nu^{2)}$ & $\Lambda\sim {\mathcal O}($ TeV)   & $\Lambda\sim {\mathcal O}(30$ TeV) & $\sim 10^{13}$ \\
$b\to ss\bar d^{3)}$ & $\Lambda\sim {\mathcal O}($ TeV)   & $\Lambda\sim {\mathcal O}(10^3$ TeV) & $\sim 10^{13}$ \\
$\beta$ from $B\to J/\psi K_S^{4)}$ & $\Lambda\sim {\mathcal O}(50$ TeV)   & $\Lambda\sim {\mathcal O}(200$ TeV) & $\sim 10^{12}$ \\
$K-\bar K$ mixing{}$^{5)}$ & $\Lambda>0.4$ TeV ($6$ TeV)   & $\Lambda>10^{3(4)}$ TeV & now \\
 \hline\hline
\end{tabular}
\caption{\sl\small The ultimate NP scales that can be probed using different observables listed in the first column. They are given by saturating the theoretical errors given respectively by 1) $\delta \gamma/\gamma=10^{-6}$, 2) optimistically assuming no error on $f_B$, so that ultimate theoretical error just from electroweak corrections,  3) using SM predictions in \cite{Pirjol:2009vz},
4) optimistically assuming perturbative error estimates $\delta \beta/\beta~0.1\%$ \cite{Grossman:2002bu}, and 5) from bounds for ${\rm Re} C_1 ({\rm Im} C_1)$ from UTfitter \cite{Bona:2007vi}.}
\label{table:probes}
\end{table}

Before reaching this level of precision, of course, many practical obstacles need to be met, both experimental and theoretical. One important question would for instance be the description of electromagnetic corrections. Ignoring these difficulties let us nevertheless dream of what one could do with such a precisely measured $\gamma$. How high are the NP scales one could probe? Assuming MFV one can probe $\Lambda \sim 10^2$ TeV, while assuming general flavor violating (FV) NP one can probe $\Lambda\sim 10^3$ TeV. The reason for such high scales is the small theoretical error on $\gamma$ from $B\to D K$. It can be potentially by far the most precise probe of (N)MFV, as shown in Table \ref{table:probes}.
The downside is the immense amount of data one would need, $\sim 10^{18}$ $B\bar B$ pairs. Also, the measurement of $\gamma$ by itself means nothing. One needs an overconstrained system in order to test SM, so for the precision test of KM mechanism one needs three independent observables (to measure, e.g. $\bar\rho$, $\bar\eta$ and another one for the test). 

In conclusion, $\gamma$ extraction from $B^\pm \to D K^\pm$ offers a theoretically very clean measurement of SM CKM phase, where irreducible error comes from electroweak corrections and is estimated to be $\delta \gamma/\gamma\sim {\mathcal O}(10^{-6})$.
\Acknowledgements
I would like to thank R. Fleischer, D. Pirjol and S. Ricciardi for careful reading of the manuscript. This work was supported in part by the EU Marie Curie  IEF Grant. no. PIEF-GA-2009-252847 and by the Slovenian Research Agency.

\end{document}